\documentclass[longauth]{aa}  

\usepackage{graphicx}
\usepackage{txfonts}
\usepackage[dvipsnames]{xcolor}
\usepackage{float}
\usepackage{latexsym}
\usepackage{amssymb}
\usepackage{amsmath}
\usepackage{amsfonts}
\usepackage{color}
\usepackage{subfig}
\usepackage{dcolumn}
\usepackage[hidelinks,colorlinks=true,linkcolor=blue,citecolor=blue]{hyperref}
\usepackage[english]{babel}
\usepackage{import}
\usepackage{natbib}
\bibpunct{(}{)}{;}{a}{}{,}

\newcommand {\hi}{H{\sc I}}

\begin{document} 

   \title{The BINGO project II:}
   \subtitle{Instrument description}

\author{
Carlos A. Wuensche
\fnmsep\thanks{ca.wuensche@inpe.br}\inst{1},
    Thyrso~Villela\inst{1,2,29},
    Elcio~Abdalla\inst{3}, 
    Vincenzo~Liccardo\inst{1},
    Frederico~Vieira\inst{1},
    Ian~Browne\inst{4},
    Michael~W.~Peel\inst{5,6},
    Christopher~Radcliffe\inst{7},
    Filipe~B.~Abdalla\inst{1,3,8,9}, 
    Alessandro~Marins\inst{3}, 
    Luciano~Barosi\inst{10}, 
    Francisco~A.~Brito\inst{10,11},
    Amilcar~R.~Queiroz\inst{10}, 
    Bin~Wang\inst{12,13}, 
    Andre~A.~Costa\inst{12},
    Elisa~G.M.~Ferreira\inst{3,14},
    Karin~S.F.~Fornazier\inst{3,8},    
    Ricardo~G.~Landim\inst{15},
    Camila~P.~Novaes\inst{1},
    Larissa~Santos\inst{12,13},
    Marcelo~V.~dos~Santos\inst{10},
    Jiajun~Zhang\inst{16}, 
    Tianyue~Chen\inst{4,17},
    Jacques~Delabrouille\inst{18,19,28},
    Clive~Dickinson\inst{4},
    Giancarlo~de~Gasperis\inst{20,21},
    Edmar~C.~Gurjão\inst{22},
    Stuart~Harper\inst{4},
    Yin-Zhe~Ma\inst{23,24},
    Telmo~Machado\inst{1},
    Bruno~Maffei\inst{25},
    Eduardo~J.~de~Mericia\inst{1},
    Christian~Monstein\inst{26},
    Pablo~Motta\inst{3},
    Carlos~H.N.~Otobone\inst{3},
    Luiz~A.~Reitano\inst{1},
    Mathieu~Remazeilles\inst{4},
    Sambit~Roychowdhury \inst{4},
    Jo\~ao~R.L.~Santos\inst{10},
    Alexandre~J.R.~Serres\inst{22},
    Andreia~P.~Souza\inst{3},
    Cesar~Strauss\inst{1},
    Jordany~Vieira\inst{3},
    Haiguang~Xu\inst{27}
}
\institute{Divis\~ao de Astrof\'isica, Instituto Nacional de Pesquisas Espaciais - INPE, Av. dos Astronautas 1758, 12227-010 - S\~ao Jos\'e dos Campos, SP, Brazil 
\email{ca.wuensche@inpe.br}
    \and 
Centro de Gest\~ao e Estudos Estrat\'egicos
SCS Qd 9, Lote C, Torre C S/N Salas 401 a 405, 70308-200 - Bras\'ilia, DF, Brazil
    \and
Instituto de F\'{i}sica, Universidade de S\~ao Paulo, R. do Mat\~ao, 1371 - Butant\~a, 05508-09 - S\~ao Paulo, SP, Brazil
    \and
Jodrell Bank Centre for Astrophysics, Department of Physics and Astronomy, The University of Manchester, Oxford Road, Manchester, M13 9PL, UK
    \and
Instituto de Astrof\'{i}sica de Canarias, 38200, La Laguna, Tenerife, Canary Islands, Spain 
        \and
Departamento de Astrof\'{i}sica, Universidad de La Laguna (ULL), 38206, La Laguna, Tenerife, Spain 
    \and
Phase2 Microwave Ltd., Unit 1a, Boulton Rd, Pin Green Ind. Est., Stevenage, SG1 4QX, UK 
        \and
University College London, Gower Street, London, WC1E 6BT, UK
    \and 
Department of Physics and Electronics, Rhodes University, PO Box 94, Grahamstown, 6140, South Africa
    \and
Unidade Acad\^emica de F\'{i}sica, Universidade Federal de Campina Grande, R. Apr\'{i}gio Veloso,  Bodocong\'o, 58429-900 - Campina Grande, PB, Brazil 
    \and
Departamento de Física, Universidade Federal da Paraíba, Caixa Postal 5008, 58051-970 João Pessoa, Paraíba, Brazil
    \and    
Center for Gravitation and Cosmology, YangZhou University, Yangzhou 224009, China
    \and 
School of Aeronautics and Astronautics, Shanghai Jiao Tong University, Shanghai 200240, China 
    \and
Max-Planck-Institut f\"ur Astrophysik, Karl-Schwarzschild-Str. 1, 85748, Garching, Germany
        \and    
Technische Universit\"at M\"unchen, Physik-Department T70, James-Franck-Strasse 1, 85748, Garching, Germany
        \and
Center for Theoretical Physics of the Universe, Institute for Basic Science (IBS), Daejeon 34126, Korea     
    \and
MIT Kavli Institute for Astrophysics and Space Research, Massachusetts Institute of Technology, 77 Massachusetts Ave, Cambridge, MA 02139, USA    
    \and
Laboratoire Astroparticule et Cosmologie (APC), CNRS/IN2P3, Universit\'e Paris Diderot, 75205 Paris Cedex 13, France 
    \and
IRFU, CEA, Universit\'e Paris-Saclay, 91191 Gif-sur-Yvette, France 
    \and
Dipartimento di Fisica, Universit\`a degli Studi di Roma Tor Vergata, Via della Ricerca Scientifica, 1 I-00133 Roma, Italy 
        \and
INFN Sez. di Roma 2, Via della Ricerca Scientifica, 1 I-00133 Roma, Italy 
        \and
Departamento de Engenharia Elétrica, Universidade Federal de Campina Grande, R. Aprígio Veloso, CEP 58429-900 - Campina Grande - PB, Brazil
    \and
School of Chemistry and Physics, University of KwaZulu-Natal, Westville Campus, Private Bag X54001, Durban 4000, South Africa
    \and
NAOC-UKZN Computational Astrophysics Centre (NUCAC), University of KwaZulu-Natal, Durban, 4000, South Africa
    \and
IAS, Bat 121, Universit\'e Paris-Saclay, 91405 Orsay Cedex, France
    \and
ETH Zürich, Institute for Particle Physics and Astrophysics, HIT J13.2, Wolfgang-Pauli-Strasse 27, 8093 Zürich, Switzerland 
        \and
School of Physics and Astronomy, Shanghai Jiao Tong University, Shanghai 200240, China 
    \and
Department of Astronomy, School of Physical Sciences, University of Science and Technology of China, Hefei, Anhui 230026
    \and
Instituto de F\'{i}sica, Universidade de Bras\'{i}lia, Campus Universit\'ario Darcy Ribeiro, 70910-900 - Bras\'{i}lia, DF, Brazil     
}

\abstract
{
The measurement of diffuse $21$-cm radiation from the hyperfine transition of neutral hydrogen (\textsc{Hi} signal) in different redshifts is an important tool for modern cosmology. However, detecting this faint signal with non-cryogenic receivers in single-dish telescopes is a challenging task. The \textbf{BINGO} (\textbf{B}aryon Acoustic Oscillations from \textbf{I}ntegrated \textbf{N}eutral \textbf{G}as \textbf{O}bservations) radio telescope is an instrument designed to detect baryonic acoustic oscillations (BAOs) in the cosmological \textsc{Hi} signal, in the redshift interval $0.127 \le z \le 0.449$.
}
{
This paper describes the BINGO radio telescope, including the current status of the optics, receiver, observational strategy, calibration, and the site. 
}
{
BINGO has been carefully designed to minimize systematics, being a transit instrument with no moving dishes and 28 horns operating in the frequency range $980 \le \nu \le 1260$\,MHz. Comprehensive laboratory tests were conducted for many of the BINGO subsystems and the prototypes of the receiver chain, horn, polarizer, magic tees, and transitions have been successfully tested between 2018--2020. The survey was designed to cover  $\sim 13\%$ of the sky, with the primary mirror pointing at declination $\delta=-15^{\circ}$. The telescope will see an instantaneous declination strip of $14.75^{\circ}$.
}
{
The results of the prototype tests closely meet those obtained during the modeling process, suggesting BINGO will perform according to our expectations. After one year of observations with a 60\% duty cycle and 28 horns, BINGO should achieve an expected sensitivity of 102\,$\mu K$ per 9.33\,MHz frequency channel, one polarization, and be able to measure the \textsc{Hi}\, power spectrum in a competitive time frame.
}
{}

\authorrunning{C.A. Wuensche et al.}
\titlerunning{The BINGO project II: Instrument Description}
\date{Received: XX/XX/2021. Accepted:XX/XX/2021}

\maketitle

\section{Introduction}
\label{intro}

The observed accelerated expansion of the Universe and its possible relation to the so-called dark energy (hereafter, DE) properties is one of the most exciting questions in $21^{\rm{st}}$ century cosmology. DE properties can be investigated through baryon acoustic oscillations (BAO), one of the most powerful probes for cosmological studies \citep{Albrecht:2006,Weinberg:2013}, detected in the analysis of large optical galaxy redshift surveys \citep{Eisenstein:2005-BAO,Anderson:2014-BAO,Planck2018:cosmoparams,Alam:2020} ($z \lesssim 3.5 $). It is important to confirm this detection in other wavebands and over a wider range of redshifts.

The BINGO (\textbf{B}aryon Acoustic Oscillations from \textbf{I}ntegrated \textbf{N}eutral \textbf{G}as \textbf{O}bservations) radio telescope is an instrument specifically designed to observe BAOs in the frequency band covering from $980$ to $1260$\,MHz and to provide new insight into the Universe at $ 0.127 < z < 0.449$ with a dedicated instrument.  The BINGO concept was presented in \cite{Battye:2013} with later updates presented in \cite{Dickinson:2014, BigotSazy:2015, Battye:2016} and \cite{ Wuensche2018}. Other papers describing aspects of the project include the project overview \citep[Paper 1]{2020_project}, optical design \citep[Paper 3]{2020_optical_design}, mission simulations \citep[Paper 4]{2020_sky_simulation}, component separation and bispectrum analysis \citep[Paper 5]{2020_component_separation}, simulations for a mock $21$-cm catalog  \citep[Paper 6]{2020_mock_simulations}, and cosmological forecasts for BINGO \citep[Paper 7]{2020_forecast}.

The BINGO Phase 1 instrument has a compact optical configuration, with a 20\,m main semi-axis primary paraboloid and a 17.8\,m main semi-axis secondary hyperboloid, which illuminates a focal plane with 28 horns.
The search for BAOs will be carried out through a intensity mapping (IM) survey over a large area (about 13\%) of the sky to produce 3D maps of the \textsc{Hi}\, distribution. Observations will span several years with different operational phases. The project design, hardware setup and observation strategy of the project also open up the possibility of the detection of transient phenomena on very short timescales ($\lesssim 1\,{\rm ms}$), such as pulsars and fast radio bursts (FRBs) \citep{LorimerHandbook:2004,Lorimer:2007,Cordes2019fast}. The companion paper I \citep{2020_project} contains a section on transient science that includes estimates for FRB detections with BINGO. Table~\ref{tab:instrument} contains the main parameters of the telescope for operation during Phase 1. Most of the subsystems already have their prototypes built and tested, while others are in final project stages or under fabrication.

BINGO is mainly funded by FAPESP, a São Paulo state funding agency, and it has received funds from other Brazilian and Chinese agencies, as well as electronic components from the UK and China. Regarding the timeline, at the time of writing this paper, the project received a positive report after a major project design review (PDR) in July 2019. The engineering project for the telescope structure and the optical project have been completed and despite some delays caused by the Covid-19 pandemic, a major advancement was made during 2021 regarding terrain prospection, roadwork, and the area cleaning.

This paper starts with this Introduction. Section~\ref{sec:motivation}, describes the BINGO concept. Section~\ref{sec:instrument} contains a brief discussion of the optical design, feed horns, and receiver. Section \ref{sec:observations} outlines the BINGO observation strategy. The calibration strategies are briefly described in Section~\ref{sec:calibration}.  The site characteristics are presented in Section~\ref{sec:site}.
Final considerations for Phase 1 as well as the discussion of the next steps for BINGO are presented in Section~\ref{sec:status}.  

\section{Motivation}
\label{sec:motivation}

While many $21$-cm intensity mapping telescopes (either under development or in operation) for BAO science are interferometers or shared-time instruments, BINGO chooses a complementary approach, based upon a single main, dish illuminating many horns equipped with stable receivers. This represents an approach to BAO IM studies which is, at the same time, cheaper and efficient \citep{Battye:2013}. 

The BINGO optical design produced beams with distortions from the circular shape $\lesssim -30$\,dB inside the main beam (FWHM $= 0.67^{\circ}$) and first side lobes below $-25$\,dB for all frequencies in the BINGO band. These requirements are important and  necessary for the accurate separation of foregrounds and $21$-cm information, as discussed in \citet{2020_optical_design}.

\begin{table}[t]
 \scriptsize
\caption{BINGO project parameters - {Phase 1}}
\label{tab:instrument}
\centering
\begin{tabular}{lrl}
\hline\hline
Parameter & Value & Unit\\
\hline
\multicolumn{3}{c}{\textbf{Receiver}} \\
T$_{\rm sys}$                               & 70            & K                  \\
Lower frequency cutoff ($f_{\rm{min}}$)     & 980           & MHz        \\
Higher frequency cutoff ($f_{\rm{max}}$)    & 1260          & MHz        \\
Frequency band (B)                          & 280           & MHz       \\
Number of redshift bands\tablefootmark{a}   & 30                &         \\
Sampling time (Hz)                          & $\tau_s$      & 10  \\
Instrument noise (1\,s, 1 redshift band)\tablefootmark{b}    & $23$          & mK    \\
\multicolumn{3}{c}{\textbf{Telescope}} \\
\multicolumn{2}{c}{Primary dish (Paraboloid)} &  \\
Semi-major axis\tablefootmark{c}            & 25.5 & m          \\
Semi-minor axis \tablefootmark{d}           & 20.0 & m             \\
Concavity                                   &  0.7 & m                  \\      
\multicolumn{2}{c}{Secondary dish (Hyperboloid)} &  \\
Semi-major axis\tablefootmark{e}            & 18.3 & m  \\
Semi-minor axis\tablefootmark{f}            & 17.8 & m  \\
Concavity                                   & 0.5  & m  \\
& & \\
Total area  ($A_{\rm tot}$)                 & 1602        & m$^2$         \\
Mean effective area ($A_{\rm eff}$)\tablefootmark{g}              & 621        & m$^2$              \\
Telescope azimuth orientation                         & 0           & deg (north)\\
Focal plane area \tablefootmark{h}          & $18.6 \times 7.8$     & m$^2$  \\
Instantaneous focal plane area\tablefootmark{i}   & $14.75 \times 6.0$     & (Dec, RA) deg. \\
\multicolumn{3}{c}{\textbf{Optics}} \\
Focal length                                & 63.2          & m  \\
Telescope beam solid angle ($\Omega_{\rm{beam}}$)\tablefootmark{j}   & 0.35  & sqr deg    \\
Optics FWHM (@1.1 GHz)                      & $\sim$ 0.67    & deg \\
Number of feed horns (Phase 1)                      & 28                &         \\
Number of circular polarizations                & 2             &       \\
Vertical horn center separation             & 2.4           & m \\
\multicolumn{3}{c}{\textbf{Survey}} \\
Maximum declination coverage\tablefootmark{k} & $15.31$        & deg.        \\
Mission duration (Phase 1)                      & 5                     & years      \\
Full survey area (5 yr)                     & 5324           & sqr. deg.   \\
Duty cycle (Phase 1)                        & 60 - 90               &\% of the time        \\
\hline

\end{tabular}
\tablefoot{\scriptsize \\
\tablefoottext{a}{Assumed for science simulations, operational value will be downsampled from the digital backend output} \\
\tablefoottext{b}{Considering 28 horns, one polarization, one redshift band, and 60\% duty cycle} \\
\tablefoottext{c, d, e, f}{computed by GRASP} \\
\tablefoottext{g}{Average from individual areas computed by GRASP. Actual values vary from 532 - 653 m$^2$} \\
\tablefoottext{h}{Measured from the edges of the array presented in Figure \ref{fig:focalplane}}   \\
\tablefoottext{i}{Measured from center to center of the horns} \\
\tablefoottext{j}{Beam solid angle value computed for $f=1120$\,MHz} \\
\tablefoottext{k}{Considering $\pm30$ cm of vertical displacement, as described in Section \ref{sec:observations}}
}
\end{table}

The main difficulty in doing a statistically significant BAO detection at radio frequencies is to properly handle and remove the high intensity Galactic and extragalactic signals in the BINGO frequency band. The mean brightness temperature of the $21$-cm line is about 200 $\mu K$ in the BINGO frequency band \citep{Hall:2013}, which is not easily detected even with cryogenic receivers. To achieve the sensitivity needed for BAO detection, the accumulated integration time per pixel ($t_{\rm pix}$) must be larger than 1 day over the BINGO Phase 1 duration, which is expected to last five years.

The BINGO telescope configuration is such that a point in the sky remains within a given horn beam
for $\approx  2.7$ minutes per day at the chosen BINGO declination. The total integration time per pixel is built by many returns to the same patch of sky over the year. The number of pixels in the resulting map is given by the following: 

\begin{equation} 
    \label{eq:pixels} 
    N = \frac{\Omega_{\rm sur}}{\Omega_{\rm beam}} \,,    
\end{equation} 

\noindent where  $\Omega_{\rm beam}$ and $\Omega_{\rm sur}$ are the beam and survey coverage areas, respectively, as given in Table \ref{tab:instrument}. The total integration time available to be distributed between the pixels is the number of beams multiplied by $t_{\rm sur}$, the duration of the survey. We can then work out the signal-to-noise per pixel. The integration time per pixel, $t_{\rm pixel}$, is given by 

\begin{equation} 
    \label{eq:integration} 
    t_{\rm pixel}=28\frac{ t_{\rm sur}}{N}\,, 
\end{equation} 

\noindent where $t_{\rm sur}$ is the survey duration. We assume the same $T_{\rm sys}$ for all receivers. Each horn will observe a single declination strip in the sky and the correlation receivers (see section \ref{sec:receiver}) will produce two data streams, one for
each polarization, for the sky signal referenced against a stable load. The combination of the two output streams should deliver a theoretical sensitivity per pixel given by the following: 

\begin{equation} 
    \label{eq:sensitivity} 
    \Delta T_{min} = \frac{T_{\rm sys}}{\sqrt{t_{\rm pixel}\Delta \nu}} \,, 
\end{equation} 

\noindent where $\Delta \nu$ and $T_{\rm sys}$ are the frequency band and the system temperature, respectively, while  $t_{\rm pixel}$ is given by equation \ref{eq:integration}. Using the parameters listed in Table~\ref{tab:instrument},  a channel bandwidth of  $\Delta \nu = 9.33$\,MHz,  1 year of observations at $60\%$ duty cycle, and  the telescope, with 28 horns, should achieve a sensitivity of around 102\,$\mu$K for one polarization. Sensitivity estimates for one horn and 28 horns (considering one polarization) for 1, 2, and 5 years of observations are presented in Table \ref{tab:sensitivity}.

The BINGO configuration allows for an improvement in sensitivity through lower $T_{\rm sys}$, a longer integration time, and increasing the  number of horns.  In addition, the survey area can be increased by moving the focal plane. This last possibility will be discussed in section~\ref{sec:optics}.

\section{Instrument description}
\label{sec:instrument}

BINGO will start Phase 1 with the telescope configuration presented in Table \ref{tab:instrument}. It will collect two hands of circularly polarized signal, each of which will be fed into magic tees fronting correlation receivers. Each magic tee sends the data to two amplifier chains and, after some electronic processing (filtering, phase switching), the signal will be recombined in a second hybrid and then sent to a digital backend for frequency decomposition.

We kept two key points in mind when determining the BINGO design and observational strategy.
The first one is simplicity: in order to minimize costs, the dishes are fixed, making BINGO a transit telescope, using the Earth rotation to define the survey area. This allows one to revisit the same patch of sky every day. A fixed pointing system for such large reflectors requires much simpler mechanical structures and, almost certainly, introduces fewer systematic effects. The choice of a transit telescope defines several parameters related to the observational strategy, such as the sky region to be observed, the level of tolerable ground pickup due to the position of the reflectors on the site, and the calibration strategies using celestial sources.

The second key point relates to the parameters defined by the optical design to optimize the focal plane, requiring small vertical and longitudinal displacements of the individual horns. This is equivalent to displacing the entire focal plane in order to increase the observed sky region.

The optical system follows an off-axis, crossed-Dragone configuration \citep{Dragone:1978}, with a primary offset paraboloid with a 25.5\,m semi-major axis and secondary offset hyperbolid with a 18.3\,m semi-major axis as reflectors. The total area of the primary is $1602~\textrm{m}^2$, while its effective area is approximately $1120~\textrm{m}^2$.  The telescope focal length  is 63.2\,m and the primary mirror points to declination $\delta=-15^{\circ}$. The illumination of the secondary by the focal plane array delivers an instantaneous field-of-view of  $88$ square degrees, with an angular resolution (FWHM) of $\approx 0.67^{\circ}$. We deliberately under-illuminate the secondary in order to reduce spillover and reduce far out sidelobes. The under-illumination is reflected in the small effective area of each horn, at central frequency $f=1100~$ MHz, shown in table \ref{tab:areas}.

\begin{table}[h]
 \scriptsize
\caption{Effective area of BINGO horn beams (computed at 1100 MHz).}
\label{tab:areas}
\centering
\begin{tabular}{|c|c|c|c|c|c|c|c|}
\hline\hline
Horn & Area    & Horn & Area    & Horn & Area    & Horn & Area  \\
     & (m$^2$) &      & (m$^2$) &      & (m$^2$) &.     & (m$^2$) \\
\hline
1  & 637.8 & 2  & 648.3 & 3  & 649.7 & 4  & 642.3 \\
5  & 626.4 & 6  & 602.7 & 7  & 572.0 & 8  & 646.2 \\
9  & 652.3 & 10 & 648.8 & 11 & 638.3 & 12 & 617.8 \\
13 & 590.9 & 14 & 554.3 & 15 & 650.4 & 16 & 652.7 \\
17 & 647.6 & 18 & 634.1 & 19 & 610.4 & 20 & 583.2 \\
21 & 531.6 & 22 & 641.3 & 23 & 648.3 & 24 & 647.8 \\
25 & 640.1 & 26 & 620.5 & 27 & 596.4 & 28 & 560.8 \\
\hline
\end{tabular}
\end{table}

We estimate the required dish surface accuracy as $\lesssim \lambda / 60 $, which, at the shortest operational wavelength of 23.8\,cm, gives a requirement of $\approx0.4$\,cm. With the combination of the two mirrors, this will give a combined surface efficiency (according to the Ruze formula) of 92\,\%, or a signal-to-noise decrease of $<10$\,\%.

Optical design simulations \citep{2020_optical_design} and measurements of the beam profile of the first prototype horn \citep{Wuensche:2020} indicate that the instrument will produce a very clean beam, with secondary lobes usually at $\lesssim -25$\,dB and very good cross-polarization rejection ($\lesssim -30$\,dB) (see Figure \ref{fig:beampatterns}). This will contribute to good performance of the component separation process  \citep{Remazeilles:2011,BigotSazy:2015,Olivari:2016,Olivari:2018,Carucci:2020}. The angular resolution of $\approx 0.67^{\circ}$ is suitable to resolve structures of an angular size corresponding to a linear scale of around $150$ Mpc in the BINGO redshift range. 

\begin{figure}[h]
    \centering
    \includegraphics[width=8cm]{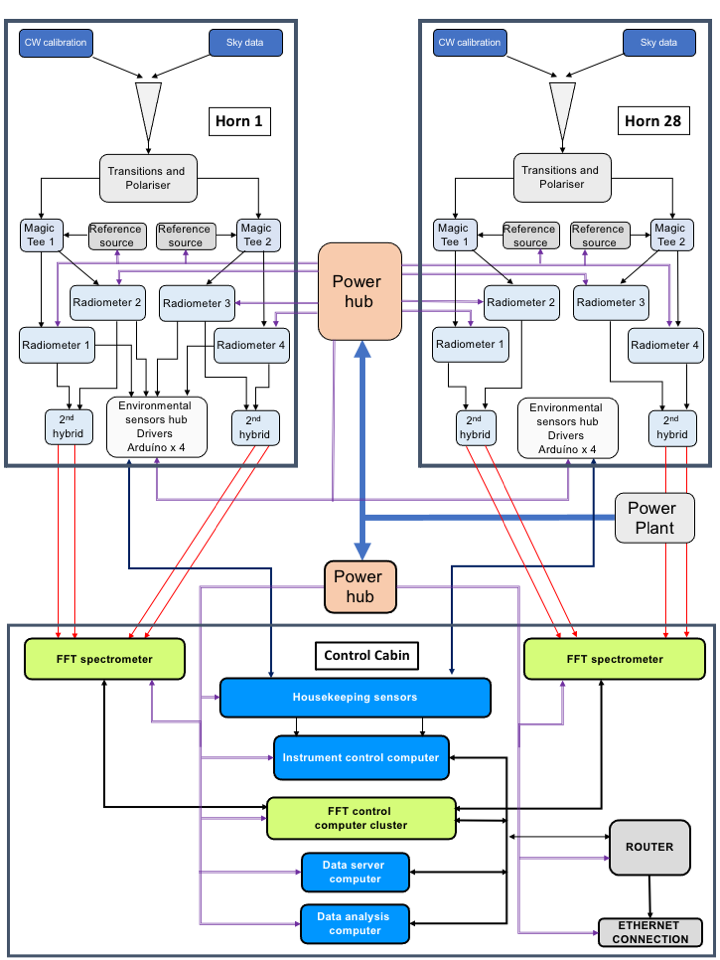}
    \caption{Top level block diagram for telescope and control cabin. }
    \label{fig:blockdiag}
\end{figure}

\begin{figure}[h]
    \centering
    \includegraphics[width=9cm]{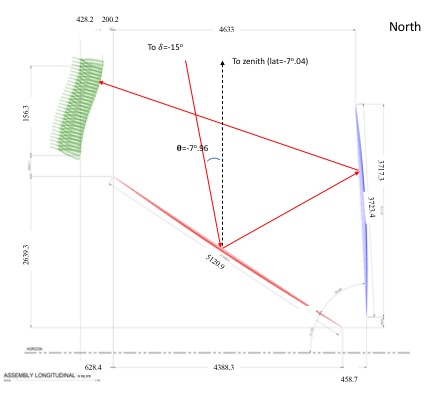}
    \caption{BINGO optics schematics: the primary mirror is in the center, facing north pointing to $\delta = -15^{\circ}$. All dimensions are in millimeters. }
    \label{fig:opticsview}  
\end{figure}

\begin{figure}[h]
    \centering
    \includegraphics[width=9cm]{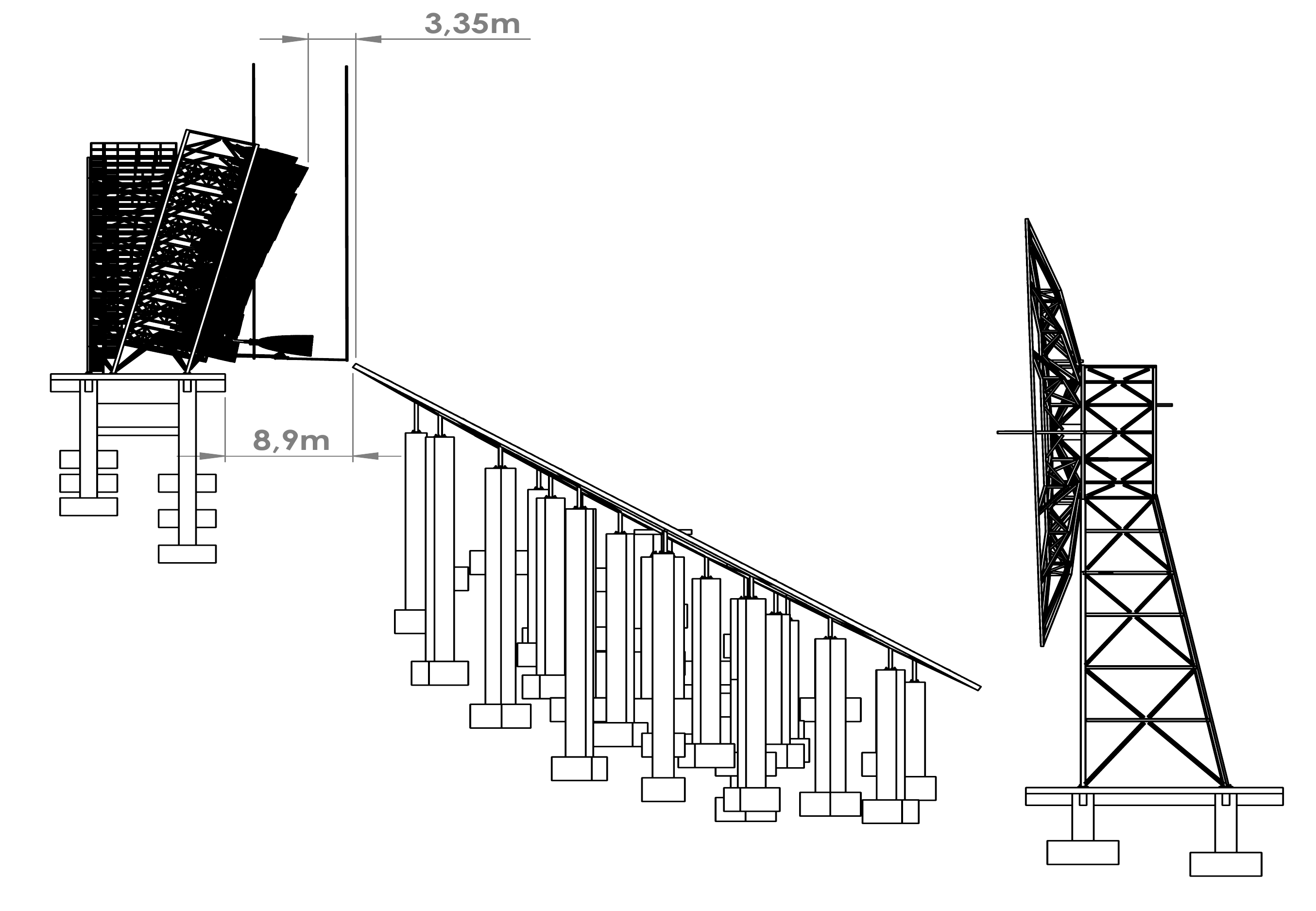}
    \caption{Engineering design based upon the dimensions and angles shown in Figure \ref{fig:opticsview}. We note that the exact positions for the support structure foundations have already been calculated.}
    \label{fig:engineeringview}    
\end{figure}

\begin{figure}[h]
    \centering
    \includegraphics[width=9cm]{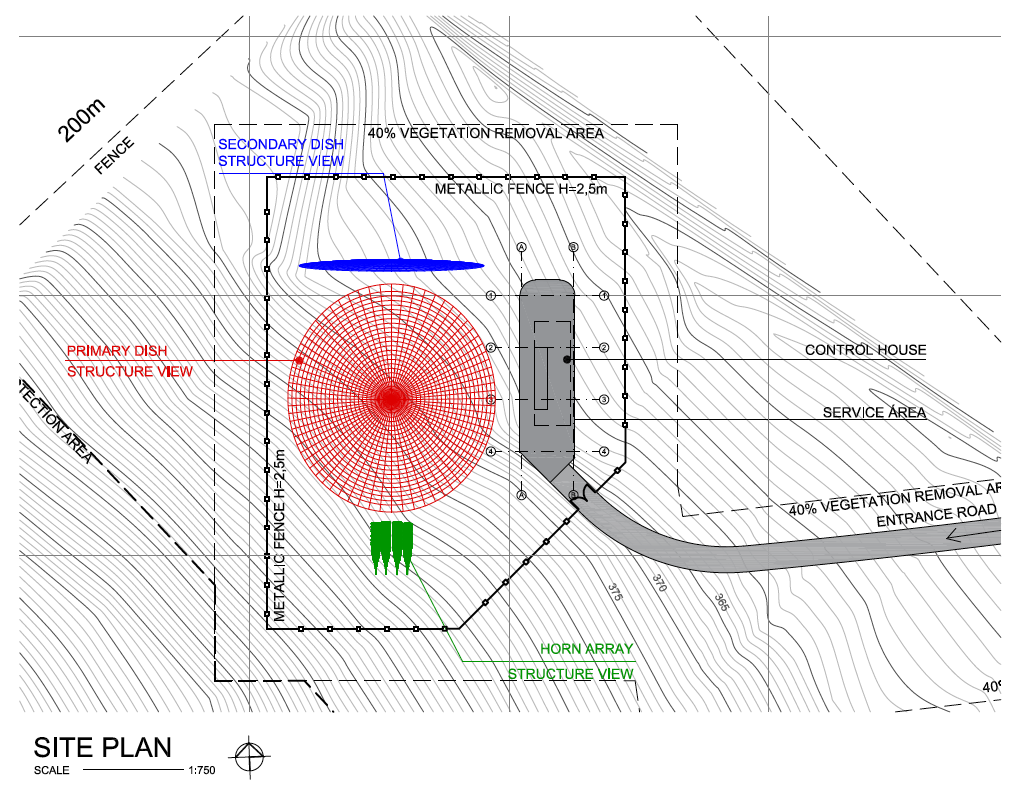}
    \caption{Plan view of the site, with the structure in Figure \ref{fig:engineeringview} highlighted in blue, red, and green. The top of the figure points to the north. The control cabin is located on the southwestern part of the terrain, behind a hill and off the main view of the telescope.}
    \label{fig:aerealview}    
\end{figure}

The top level system diagram is shown in Figure \ref{fig:blockdiag}. An overview of the optical design, engineering project for the complete telescope, as well as its location on the site, are shown in Figures~\ref{fig:opticsview}, \ref{fig:engineeringview} and \ref{fig:aerealview}. Figure \ref{fig:aerealview} also includes the topographic curves in the terrain, indicating that the shielding by the hill on the west side of the terrain is helpful to reduce radio frequency interference (hereafter, RFI) contamination. Site details are described in Section \ref{sec:site}.

We believe that the BINGO concept makes the design, instrument modeling, construction, and operation considerably simpler than that of a conventional telescope. The following sections describe the main subsystems of the BINGO telescope.

\subsection{Optics}
\label{sec:optics}

Studies for optical design and focal plane arrangement for Phase~1 were performed using the GRASP package from the TICRA Foundation\footnote{\url{http://www.ticra.com/software/grasp/}}. The BINGO instantaneous field-of-view will cover $14.75^{\circ} ~(\textrm{DEC}) \times 6.0^{\circ} ~(\textrm{RA})$, delivering an instantaneous $14.75^{\circ}$-wide declination strip, covered by 28 horns distributed in the focal plane.

\begin{figure}[h]
    \centering
    \includegraphics[height=10cm]{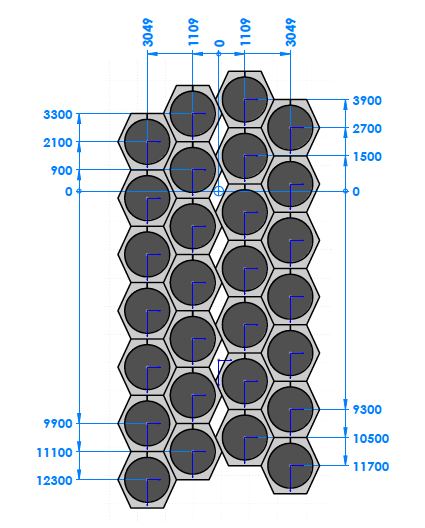}
    \caption{Horn arrangement in the focal plane for BINGO Phase 1. All dimensions are in millimeters.
    }
    \label{fig:horn_array}
\end{figure}

\begin{figure}[h]
    \centering
    \includegraphics[width=9cm]{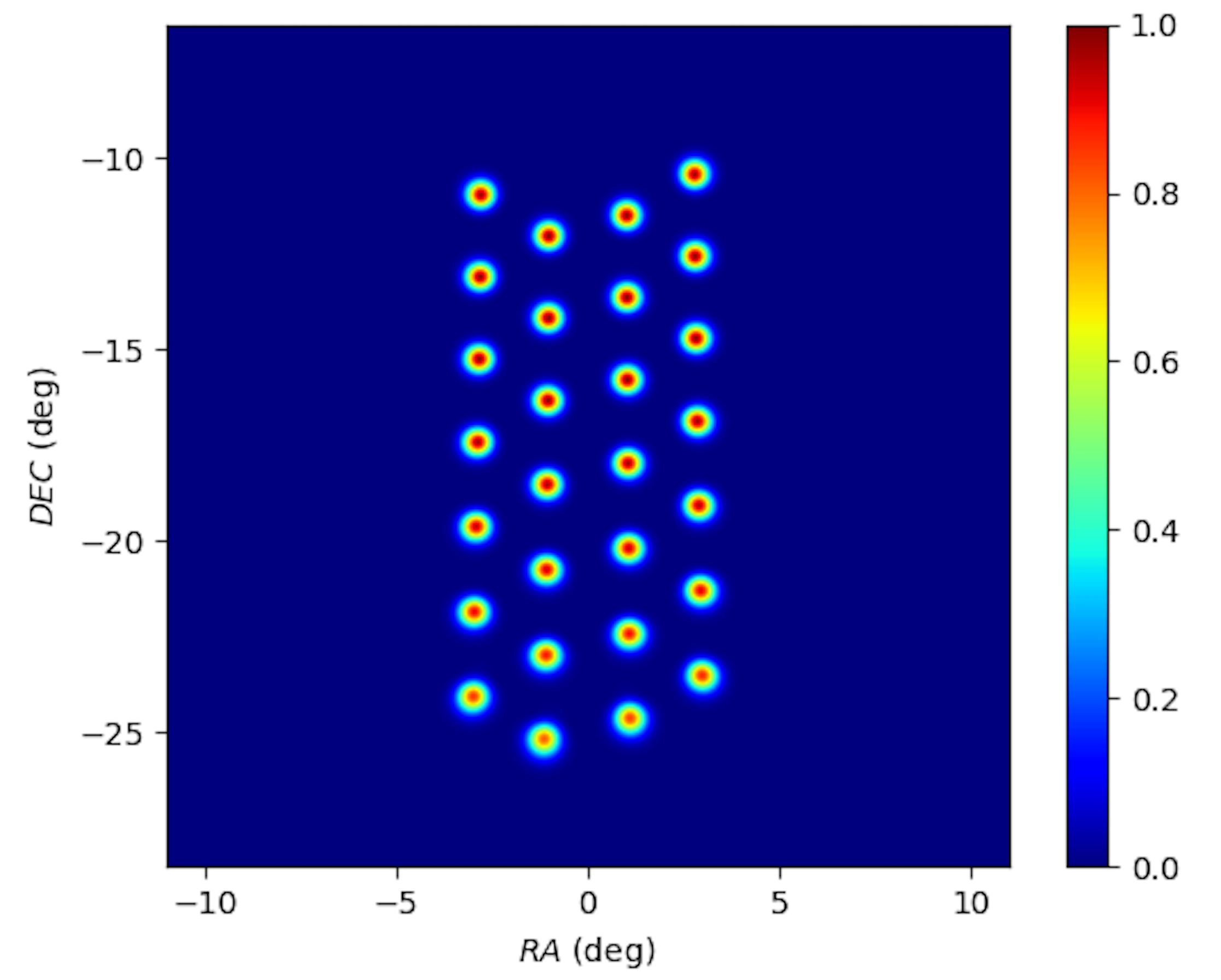}
    \caption{Beam pattern response in the focal plane, as looking at the sky in equatorial coordinates. The color scale on the right is the total intensity normalized to the peak amplitude. }
    \label{fig:focalplane}    
\end{figure}

The forward gain of each beam formed by the combination of the feed horn and the reflector does not vary by more than 1\,dB for all pixels across the focal plane. The final design guarantees both low optical aberration and very low beam ellipticity, even for beams on the edge of the focal plane. The distribution of horns in the focal plane guarantees that each horn will observe an independent patch of the sky in declination, according to the distances shown in Figure \ref{fig:horn_array}. This distribution corresponds to the optimal beam pattern configuration shown in Figure \ref{fig:focalplane}, and to the best solution for the horn arrangement in terms of uniformity of the sky coverage. We note the inverted positioning in respect to Figure \ref{fig:horn_array}, due to the way it maps the sky.

The horn pointing to the secondary reflector is defined by six parameters: the Cartesian ($x,y$) coordinates of the horn in the focal plane, the coordinates ($\theta,\phi$), corresponding to $(AZ,EL)$ positioning, and a coordinate $z$ that describes the longitudinal displacement of the horn. The pair ($x,y$) is defined by the position of the horn in the focal plane, as shown in Figure \ref{fig:horn_array}. Additionally, each horn can be displaced vertically ($\delta y$) by $\pm 300$ mm in relation to the longitudinal direction that connects the center of the hexagon to the surface of the secondary reflector. The vertical displacement $\delta y$ in $\pm 300$ corresponds to an increase of $\approx 0.56^{\circ}$ in the declination coverage, enlarging it to $-10.17^{\circ} \le \delta \le -25.48^{\circ}$ after 5 years of operation.

Each  horn can also be displaced longitudinally by $700 - 1300$ mm from the nominal zero-point where the horn opening aligns with the supporting structure, and tilted by $\pm 5^{\circ}$ in the $\phi$ (azimuth) direction. The longitudinal positioning is done by means of a long screw with a narrow thread, which allows for a precision better than 5 mm.

\begin{figure}[h]
    \centering
    \includegraphics[width=9cm]{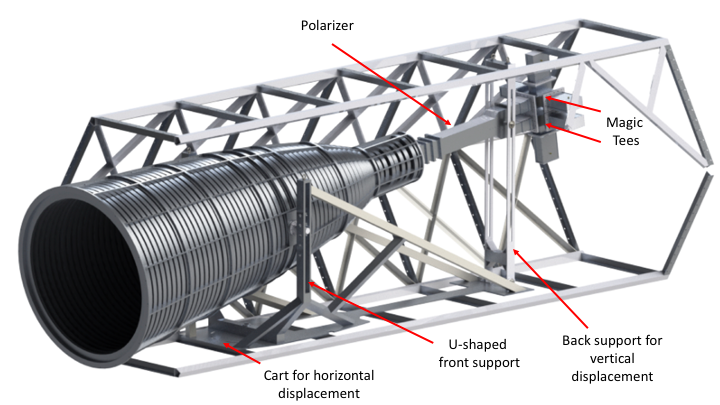}
    \caption{Hexagonal cage for the horn with dimensions 6500\,mm(length) x 2400\,mm(height) x 2600\,mm(width). The horn is attached to a U-shaped support by the ring crossing its center of mass, for pivoting and vertical displacement, and to another structure connected to the end of the polarizer, on the opposite side of the horn mouth. The U-shaped support and the back structure are mounted on top of a cart that allows for longitudinal positioning of the horn.}
    \label{fig:hexagon}
\end{figure}

Individual horns will be positioned inside hexagonal cages. A cart and a V-shaped support inside the hexagon were built to allow more precise positioning of $(\theta, \phi, z, \delta y)$ in the focal plane. Figure \ref{fig:hexagon} depicts the horn inside the hexagon, with the V-shaped support and the cart. The final positions of the optical system coordinates are defined in relation to the line connecting the center of the hexagon to the surface of the secondary reflector. Stacking up the hexagons will build the focal plane shown in Figure \ref{fig:horn_array}.

The consequence of this choice for positioning the horns is that, when the optical axis is parallel to the longitudinal direction, there will be a $2.4$-m separation between horn centers in the vertical direction. Section \ref{sec:observations} pertains to the effects of the vertical displacements in more detail.

The vertical displacement of the horns change the secondary dish illumination and it was designed to eliminate the gaps in the sky coverage that would occur if horns were kept in the same position during the full mission, leading to a less-than-optimal sampling of the sky, with a Nyquist sampling frequency $f_\mathrm{Nyq} < 1$. The combination of these adjustment parameters allows for a very fine pointing, with the horns located in the outermost regions of the array being slightly tilted with respect to the focal plane to properly illuminate the secondary reflector and reduce side lobes and ground spillover. 

A complete description of the optical design is presented in the BINGO project companion paper III \citep{2020_optical_design}. The results presented there, supported by the results of the beam profile measurements of the prototype horn \citep{Wuensche:2020}, suggest that the desired clean beam, with low cross-polarization levels as well as small beam distortion and spillover, will be achieved during operation.  

\subsection{Feed horns and front end} 
\label{sec:horns}

BINGO uses specially designed conical corrugated horns to illuminate the secondary mirror of the telescope. Because of the large focal ratio needed to provide the required wide field of view, the horns have $\thicksim 1.7$ m diameter and $\thicksim 4.9$ m length. The horn was designed to produce optimum illumination of the secondary mirror while giving excellent insertion and return loss properties. The corrugations consist of a series of rings with a width (or thickness) of 3.4\,cm, corresponding to one-eighth of the wavelength of the central observing frequency of 1.1\,GHz. The rings have alternating diameters to create the corrugation depth, and the horn should reach an edge taper below $-20$ dB at $19^\circ$ across the whole band (horn beam power at the edge of the secondary mirror of the telescope). The BINGO team worked together with local companies in the development of the horn and front end (transitions, polarizer, and magic tees) prototypes, all of which were successfully tested in the Division of Astrophysics and the Laboratory of Integration and Tests (LIT), located at Instituto Nacional de Pesquisas Espaciais (INPE).

The test range consists of a 9 m wide $\times$ 9 m tall $\times$ 10 m deep chamber, fully covered inside with absorbing cones, that faces a transmitter mounted on a tower located 80 m away from the chamber that produces both horizontal and vertical, linearly polarized signals. The system is capable of continuous $360^{\circ}$ horizontal rotation, clockwise and counterclockwise, and the horn response to the signal emitted by the far-field transmitted is continuously measured until a $360^{\circ}$ horizontal rotation is completed. The transmitter signal was generated by a ANRITSU MG3649B equipment and the horn was connected to a MI-1797 receiver fabricated by Scientific Devices Australia Pty. Ltd. - SDA. 

\begin{figure}[h]
    \centering
    \includegraphics[width=9cm]{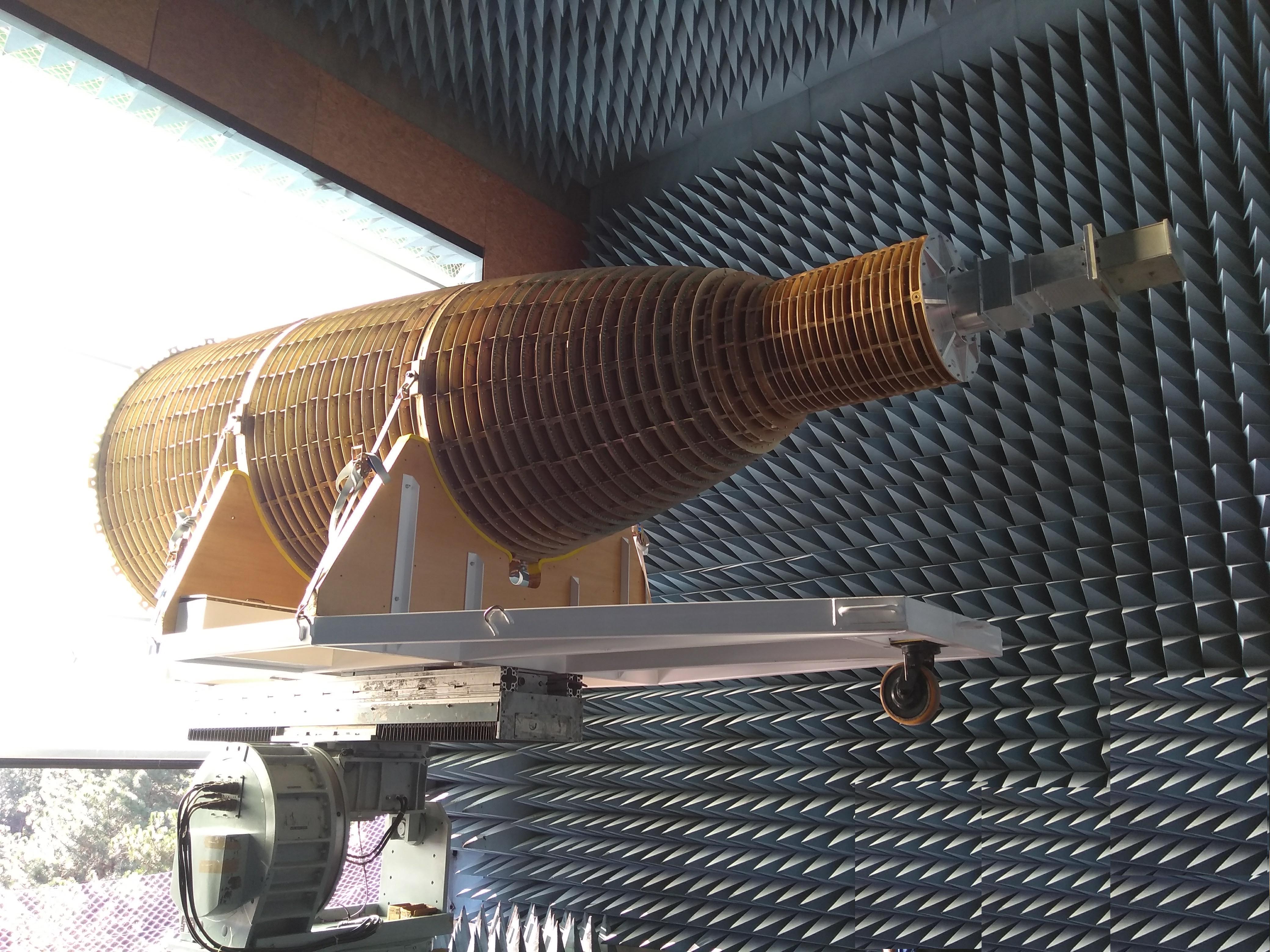}
    \caption{BINGO prototype horn in the LIT antenna test facility, ready for beam pattern measurements.} 
    \label{fig:hornLIT}
\end{figure}

We made 17 measurements at 25 MHz intervals from 900 to 1300 MHz; each frequency was measured in both vertical and horizontal direct polarization, with simultaneous cross-polarization measurements. The directivity was calculated by numerically integrating the irradiation diagrams. Figure \ref{fig:hornLIT} shows the horn and the front inside the antenna test range, during the beam pattern and polarization performance testing.

\begin{figure}[h]
    \centering
    \includegraphics[width=8cm]{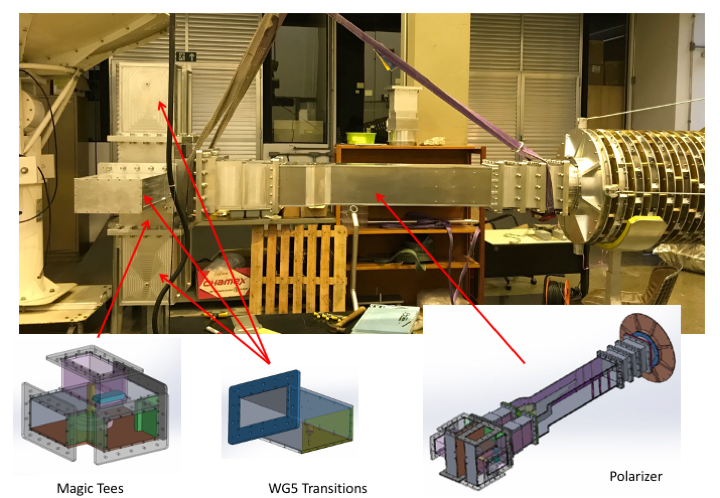}
    \caption{Front end, with parts indicated at the bottom of the figure, assembled in the throat of the horn in the integration hall at the INPE Division of Astrophysics.} 
    \label{fig:horn_backend}
\end{figure}

The complete front end, assembled in the back of the horn, during testing procedures at INPE's Division of Astrophysics, is pictured in Figure \ref{fig:horn_backend}. The insertion loss results for the measurements of the horn, polarizer, and WG5 transition prototypes are $-$0.14, $-$0.12, and $-$0.07 dB, respectively. Likewise, for the return loss, we obtained $-$26.5, $-$24.1, and $-$30\,dB.

\begin{figure*}[h]
    \centering
    \begin{minipage}{0.45\textwidth}
        \includegraphics[ height=13cm]{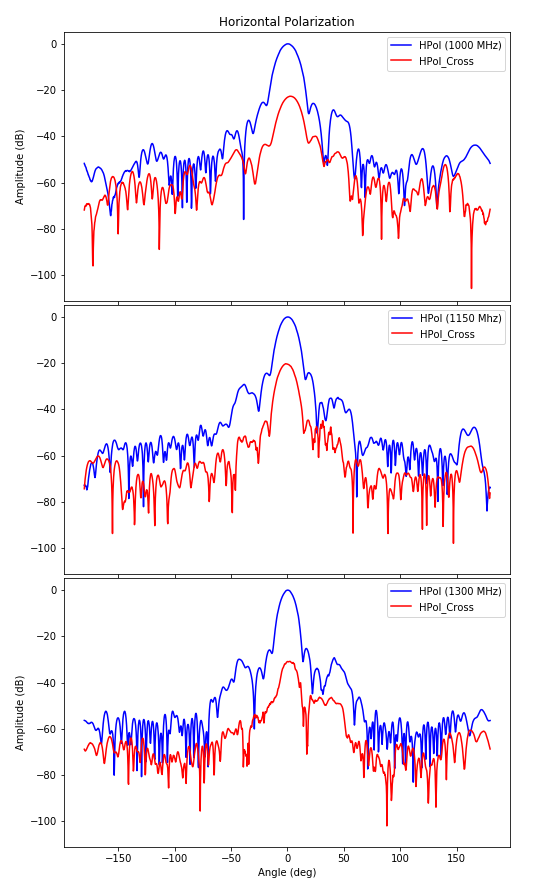}
    \end{minipage}
    \begin{minipage}{0.45\textwidth}
        \includegraphics[ height=13cm]{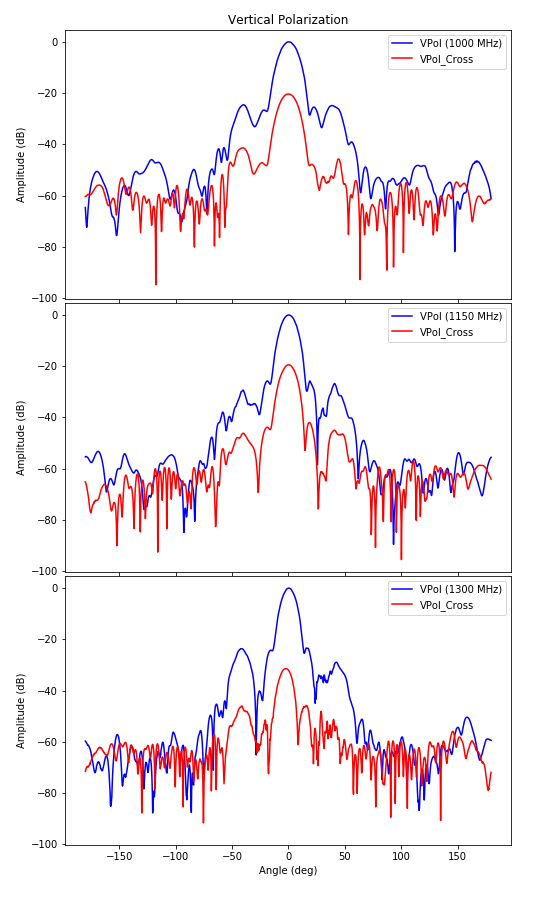}
    \end{minipage}
    \caption{Beam pattern measurements of the BINGO prototype horn. Left: Horizontal co-polarization (blue) and cross-polarization (red). We note the more than 20 dB difference between the co- and cross-polarizations in the main lobe. Right: Same as left column, but for vertical polarization. } 
    \label{fig:beampatterns}
\end{figure*}

A sample of the results of those measurements, for $1000, 1150$, and $1300$\,MHz, are shown in Figure \ref{fig:beampatterns} for both V and H (direct and cross) polarizations. The horn construction and testing, including details of the beam measurement setup, and described in detail in \cite{Wuensche:2020}. These measurements were used as input data for the simulations presented in the optical design in Paper III \citep{2020_optical_design}.

\subsection{Receiver} 
\label{sec:receiver}

The first stage filters define the BINGO operational frequency band of $980 - 1260$\,MHz, in which BINGO will operate. It is located inside a band dedicated to aeronavigation, as defined by the International Telecommunication Union (ITU). The lowest frequency limit is slightly above the 3G mobile phone band at $956$\,MHz. However, the band is prone to contamination from ADS-B airplane transponders ($1090$\,MHz) and  harmonics from geostationary satellites \citep{Harper:2018, Peel:2019}, which will surely present a challenge during the data cleaning process.

The existence of the potential problems described above means that first stage amplifiers should have a large dynamic range in order to avoid being driven nonlinearly by strong, out of band, RFI. The first stage amplifiers are followed by specially designed bandpass filters to have a very steep cutoff at the lower end of the band in order to mitigate the effects of mobile phone signal at $956$\,MHz.

We tested different radiometer chains, with MiniCircuits (\mbox{NF = 0.38\,dB}, \mbox{G = 17\,dB}), customized Skyworks (\mbox{NF = 0.35\,dB}, \mbox{G = 21\,dB}), and WantCom (\mbox{NF = 0.35\,dB}, \mbox{G = 38\,dB}) as low noise amplifiers (LNA), Phase2 Microwave filters, and Quietec isolators. Preliminary, and as of yet unpublished, measurements at Jodrell Bank Observatory of the $1^{st}$-stage Skyworks LNA indicate a receiver temperature $T_{\rm rec} = (24.9 \pm 2.0)$ K, which is considered our nominal temperature. Table \ref{tab:comp_antena} shows the temperature for each component of the BINGO radio telescope, allowing an initial estimate for the system temperature to be $T_{\rm sys} = T_A + T_{\rm{rec}} \approx 70$ K, where $T_A$ is the contribution of all components before the LNA and $T_{\rm{rec}}$. The characterization of the receiver system temperature can be used to estimate the fundamental white-noise (thermal noise) limit of the time ordered data (TOD) and is described by equation~\ref{eq:sensitivity}.

\begin{table}[h]
\centering
\caption{BINGO estimated temperature contributions to $T_{\rm sys}$.}
\label{tab:comp_antena}
\begin{tabular}{|l|c|}
\hline
\textbf{Signal contributions} & \textbf{T (K)} \\
\hline
CMB                                 & 2.7 \\
Galactic/Extragalactic signals      & $\sim 5.0$ \\
Ground spillover                    & 2.0 \\
Atmosphere                          & 4.5 \\
Parabolic mirror                    & 1.5 \\
Hyperbolic mirror                   & 1.5 \\
Horn (measured)                     & 6.8 \\
Polarizer (measured)                & 9.1 \\
Magic tee (measured)                & 2.8 \\
Rectangular-to-coax waveguide (measured)            & 9.1 \\
LNA (phase matched)                     & 24.9 \\
\hline
Total & $\sim 69.9$ \\
\hline
\end{tabular}
\end{table}

As well as system temperature variations, the BINGO receiver outputs are subject to system gain variations. For all radiometers, the gain fluctuations will be correlated over some
timescale. 
The TOD contains the voltage from the outputs of the radiometers and presents a power spectrum distribution that can be approximately modeled by a reciprocal power-law. 
This type of correlated noise is known collectively as $1/f$ noise \citep{Maino99}. In the power spectrum, the frequency at which the contributions of thermal and $1/f$ noise are equal is known as the knee frequency.  

\begin{figure}[h!]
    \centering
    \includegraphics[width=9cm]{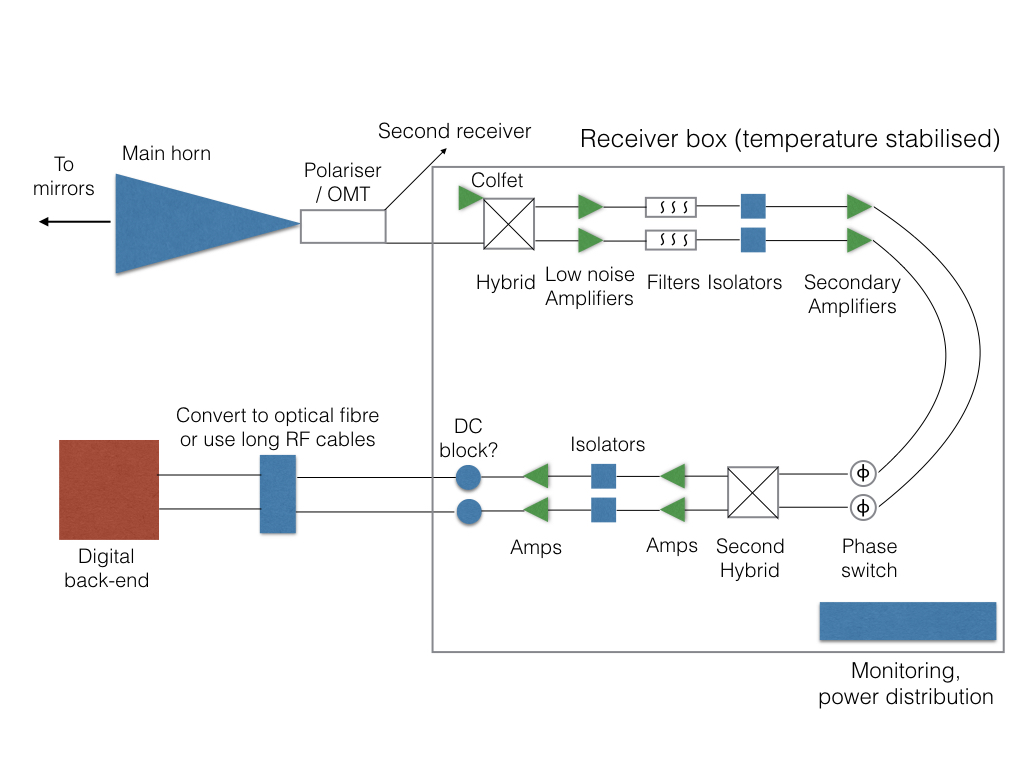}
    \caption{Block diagram of the BINGO correlation receiver. The correlation is achieved by means of the hybrids, which in our case will be waveguide magic tees (first hybrid) and striplines (second hybrid).}
    \label{fig:full_receiver}
\end{figure}

A full correlation receiver (FCR) configuration was our main receiver choice and was also recommended by the PDR committee in July 2019. The FCR is expected to remove most of the gain variations, leaving a  knee frequency $f_{knee} \sim 1~\textrm{mHz}$ and 
minimizing the $1/f$ noise contribution to the TODs \citep{Heron2011,Bersanelli10}. This configuration is schematically shown in Figure \ref{fig:full_receiver}.

Using the parameters from Table \ref{tab:instrument} in Equation \ref{eq:sensitivity}, we obtain, for one $9.33$\,MHz channel and one second integration, $ \Delta T_ {\rm min} \approx 23$ mK. Table~\ref{tab:sensitivity} contains our estimates of the optimal sensitivities for one and 28 horns (considering only one polarization), with a mission duration of 1, 2, and 5 years, and duty cycles of 60\% and 90\%, using data from Table \ref{tab:instrument}. The above estimates do not take gain variations in the receiver into account, which will degrade sensitivity.

We can use the sensitivities in Table~\ref{tab:sensitivity} to estimate the minimum flux density detected by the BINGO receiver using the expression \citep[see e.g.,][]{Kraus:1986,Wilson2013}

\begin{equation}
    S(\Delta T_{\rm min}) =  \frac{2k}{A_{\rm eff}} \Delta T_{\rm min},
    \label{eq:flux} 
\end{equation}

\noindent where $A_{\rm eff}$ is the effective area and $k$ is the Boltzmann constant.  For a single $9.33$\,MHz channel and 1 second integration, we obtain a flux $ S (\Delta T_{\rm{min}}) \approx  0.06$ Jy.

\section{Observation strategy}
\label{sec:observations}

This section describes the observation strategy used in the simulations that emulate the BINGO mission with the parameters defined for BINGO Phase 1. The chosen horn array favors the most uniform sky coverage in the declination band $-10^{\circ} \lesssim \delta \lesssim -25^{\circ}$. The focal plane covers an approximate instantaneous area of $88.5$ square degrees, allowing for a daily sky coverage of $5324$ square degrees.

\begin{table}[h]
    \centering
    \caption{BINGO estimated sensitivity for a 60 (90) \% duty cycle (one polarization).}
    \label{tab:sensitivity}
    \begin{tabular}{|c|c|c|c|} 
    \hline
            & \multicolumn{3}{c|} { $\Delta T_{\rm min}$ ($\mu {\rm K}$)} \\
    \hline
    \ Number of horns & 1 year & 2 years & 5 years \\
    \hline
     1  &  542 (442)&  383 (313)& 242 (198) \\
     28 &  102 (84)  & 72 (59)   & 46 (37)      \\    
    \hline
    \end{tabular}
\end{table}

The BINGO optical design contemplates an increased coverage in declination by vertically moving all horns in the focal plane at once in steps of 150mm. Five different horn positions are allowed: the "zero" position (in a reference frame of the hexagon), two positions at  $\pm$ 150 mm, and two positions at  $\pm$ 300 mm. All $\pm$ displacements refer to the "zero" position. The Phase 1 plan is to change the horn positions in the focal plane every year, covering one position per year. 

After five years of observations, with the focal plane displaced by  $\pm 150\textrm{mm}$ and $\pm 300\textrm{mm}$ in the y-direction, the declination coverage will be $-10^{\circ} 10'.4 \le \delta \le 25^{\circ} 28'.93$. This declination range overlaps with a few interesting optical surveys, such as DES, 6dFGRS, Pan-Starss, and the upcoming LSST, and decreases the contribution to the data from GNSS transmitters, located at lower declinations. 

\begin{figure}[ht]
    \centering
    \includegraphics[width=9cm]{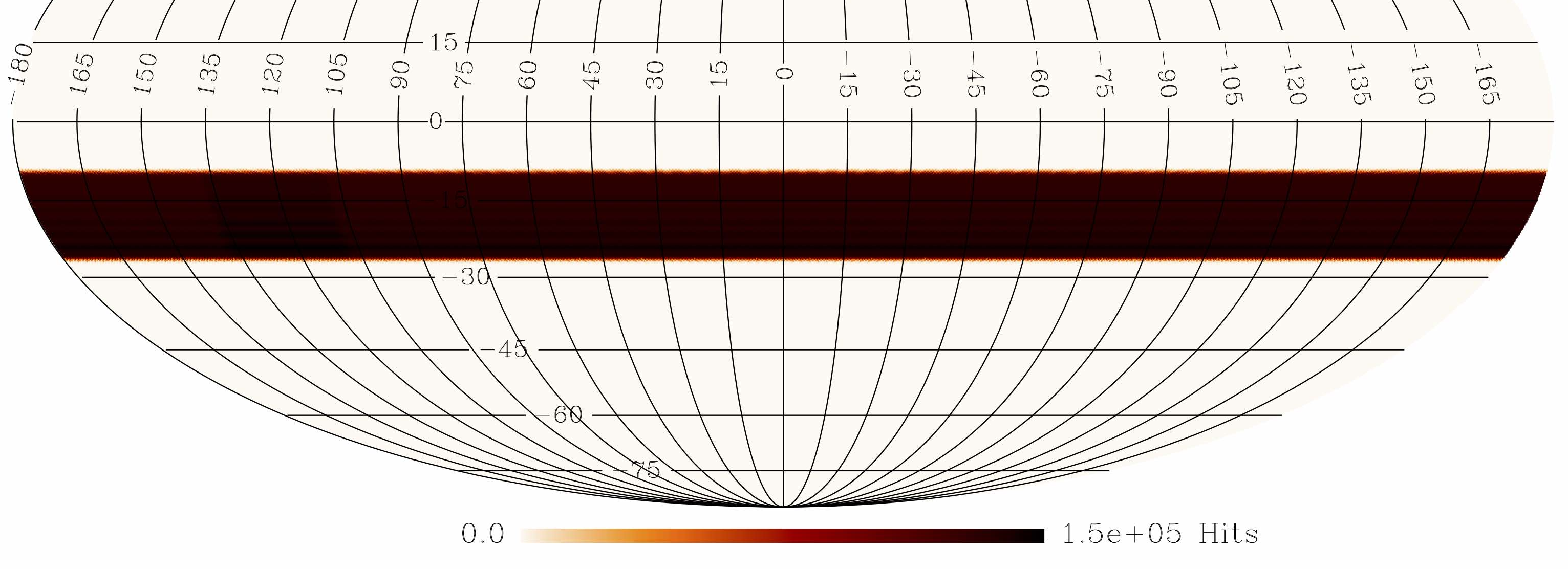}
    \caption{Resulting BINGO coverage (hits map) for one year of observations with the double-rectangular array. In this figure, the 28 feed horns are in "zero position," as explained in the text.}    \label{fig:coverage}
\end{figure}

Figure \ref{fig:coverage} shows a fraction of the sky coverage, centered in dec$=-15^{\circ}$, when using the Phase~1 nominal focal plane for a 5-year mission simulation. One can clearly see the uniform coverage delivered by this configuration, except for the pixels at the border, which are undersampled. The 28 horns in the focal plane cover the complete right ascension range at a constant, different declination for each horn, with a very small beam overlap for adjacent horns.

\begin{figure}[h]
    \centering
    \includegraphics[width=9cm]{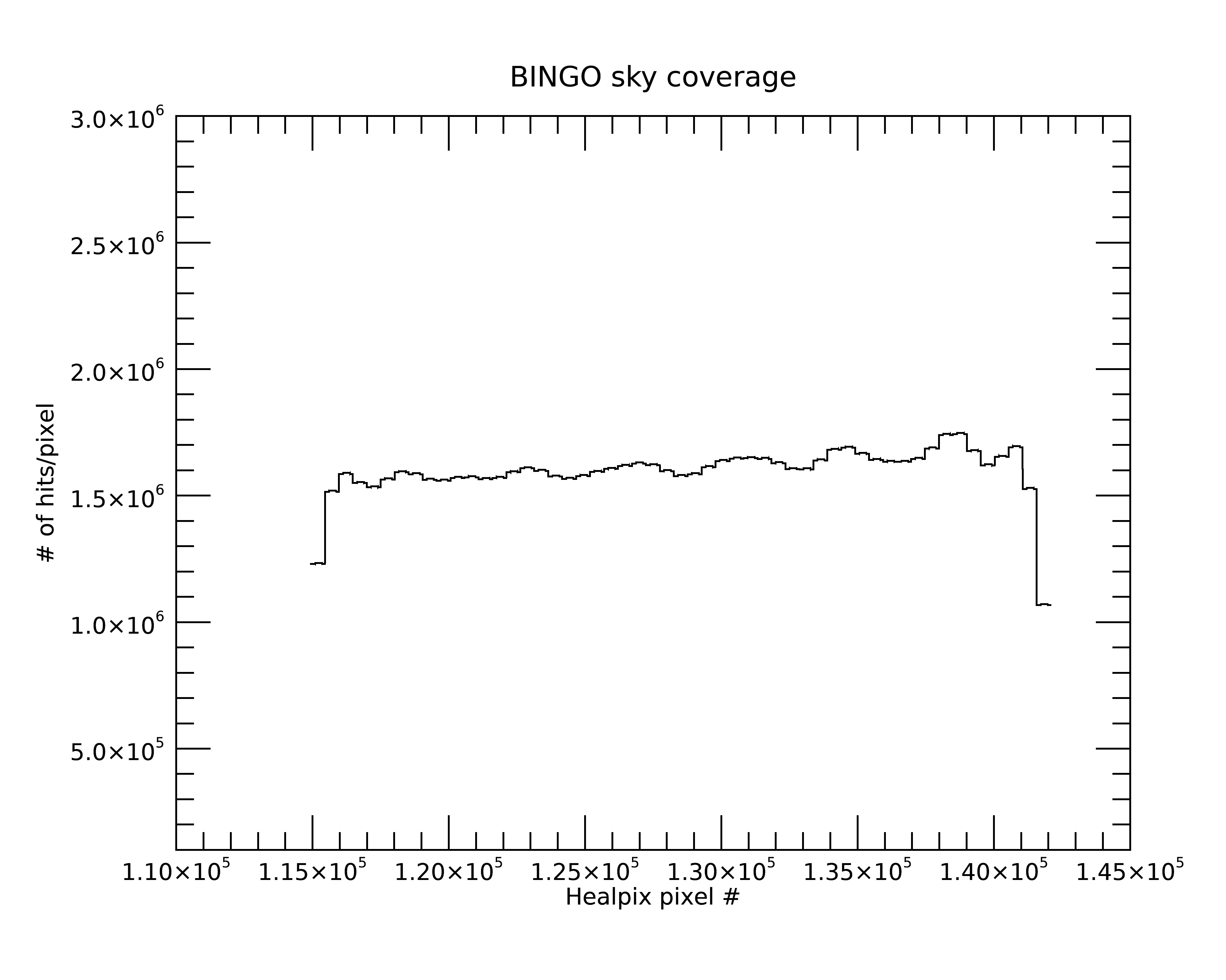}
    \caption{Histogram of hits per pixel for the above sky coverage after a mission simulation of 5 years (Healpix NSIDE=128). We note the almost uniform coverage per pixel across the BINGO band.}
    \label{fig:histogram}
\end{figure}

The result of this observation strategy can also be seen in Figure \ref{fig:histogram}, where the minimum number of observations per HEALPix\footnote{\textbf{H}ierarchical \textbf{E}qual \textbf{A}rea iso\textbf{L}atitude \textbf{Pix}elation. \citep{Healpix:2005}, \url{https://healpix.sourceforge.io}} pixel (hits) as a function of the covered area is shown. The strategy of allowing the horns to be displaced in the vertical direction gives  results in a more uniform covered area (12\% more Healpix pixels when compared to the pixels covered in the "zero" position only), which will allow us to extract the \textsc{Hi} power spectrum more efficiently. Furthermore, Phase 1 configuration allows for a sufficient Nyquist sampling per beam width ($\sim$ 1.2) during observing campaigns. The complete description of the simulations for the BINGO mission is found in  companion paper IV (\citealp{2020_sky_simulation}).

\section{Calibration}
\label{sec:calibration}

There are several facets to the instrumental calibration and not all of these can be addressed with a single approach. For clarity, we identify the individual calibration challenges and set out how we plan to address them. Though some issues can be dealt with during the data processing, it is important to minimize them as much as possible using hardware so as to simplify the processing effort.

\subsection{Gain stability}
Finding a way to achieve the required gain stability is of fundamental importance as it dictates the architecture to be adopted for each of the receivers and may limit our ability to extract the weak \textsc{Hi} signal.  Since the telescope has a fixed scan strategy (drift scan), we must rely on the gain of each receiver channel remaining constant while structures with the angular scales of interest drift through the telescope beam. For BINGO, the largest angular scales have several degrees and such structures will take tens of minutes to drift through the beams. Ideally we wish to be limited by thermal noise on these timescales and not changes in receiver gains. Conventionally, gain fluctuations have a $1/f$ power spectral slope and white noise has a zero slope. 

The impact of $1/f$ noise on intensity mapping has been explored in detail by \citet{Harper:2018b}. One of their results indicates it is important that any gain fluctuations are highly correlated across the frequency band of interest. The spectral index of the frequency correlations is defined in the limits 0 < $\beta$ < 1, where $\beta$ = 0 describes $1/f$ noise fluctuations that are identical in every frequency channel, whereas $\beta$ = 1 would describe $1/f$ noise that is independent in every channel. Preliminary measurements using a prototype of the BINGO correlation receiver at Jodrell Bank Observatory showed that the gain fluctuations did not differ much in adjacent channels, yielding an estimate for $\beta$ $\approx$ 0.25 across the full 280 MHz BINGO band, which means that $1/f$ gain fluctuations can be removed in later stages of the data analysis.

We plan to get as close to our goal of a $1$\,mHz knee frequency  as possible by using correlation receivers that take the difference between the sky signal and that from a stable reference load such as 
colfets, which are LNAs used in reverse \citep{Frater:1981}.  Preliminary laboratory measurements \citep{Evans2017} of the performance of the FCR indicate that knee frequencies of $\sim 10$\,mHz can be achieved using colfets. We are planning more tests using different types of LNA to reach the $f_{knee} \sim 1~\textrm{mHz}$ goal and the expectation is that we can achieve it with our current hardware setup.

Another promising technique for improving gain stability is the use of a single frequency, continuous wave (CW) signal generated by an oscillator \citep{Pollak:2019}. We have done some initial laboratory tests on this method and obtain similar results to those reported by \citet{Pollak:2019}. Such a technique has several advantages. A single CW can be continuously transmitted from the center of the secondary dish so that all the horns can be simultaneously illuminated. The CW signal will appear in single channels of each of the digital backends and thus one source of CW can  improve the gain stability of all receivers. We will use SKARABs\footnote{Square Kilometre Array Reconfigurable Application Board - \copyright Peralex Electronics} as the BINGO digital backends, with nominal capability of processing 16384 FFT channels. Our choice of channels will be upwards of 1000s, so sacrificing one for calibration purposes will have a negligible effect on the science outputs. 

Since measurements indicate that gain changes are coherent across the band transmission of a single frequency, 
one CW frequency channel should be sufficient, though more than one would give some redundancy.

At the same time, CW transmission  would enable the calibration of the relative gains of each horn-receiver combination, which we address in the next subsection.

\subsection{Relative gains of all the receiver channels}
In order to combine the information from all the receiver outputs from different horns and with different hands of circular polarization, it is necessary to measure their relative gains. We intend to do this in two ways: with a CW signal as described above, and also with the periodic transmission of a broadband noise diode signal, which is also from the vertex of the secondary dish. The latter, being a broadband signal, will help with bandpass calibration, day to day monitoring of the system stability, and provide an additional way of normalizing gains. Efforts will be made to make the frequency response of the transmitting antenna as smooth as possible. Having an overall flat bandpass is helped by the cross-Dragone optical design of the telescope, one of whose nice features is that it has minimal standing waves.

\subsection{System temperature fluctuations}
Being additive, system temperature fluctuations are indistinguishable from sky temperature fluctuations picked up by each horn antenna. Therefore, system temperature fluctuations represent a serious problem that conventional calibration techniques are unable to address. Even if the contribution of the receiver electronics to the system is perfectly stable, any attenuation in front of the amplifiers will contribute to the system noise and any change in the physical temperature of these lossy components  will change the overall system temperature. The effect of ambient temperature fluctuations on the horn and other waveguide components is of most concern. The obvious ways to mitigate the problem are: a) minimize the loss in all waveguide components; b)  try to keep them all at the same physical temperature; and c) maximize the thermal inertia of these components. Thermal inertia will damp out fluctuations on a short timescale. We will deploy all of these measures.  In terms of calibration, we will monitor the physical temperature of components and use the measured values to derive correction factors for the system temperature.

A rough estimate of the requirements on physical temperature stability can be made using the numbers quoted in Table \ref{tab:comp_antena}. 
Our goal is to produce a receiver whose stability  is limited by thermal noise for at least 100\,s.
Using equation \ref{eq:sensitivity}, $\Delta T/T$ for 100 s and a bandwidth of $280$\,MHz the thermal noise fluctuations will be $6.4 \times 10^{-6}$. We require the changes in the system temperature arising from the lossy components to be of this order or less. The horn and waveguide components together contribute approximately $28$\,K to the overall system temperature. For simplicity we assume that all components share the same physical temperature. 

Changing this temperature by 1 K from an ambient temperature $T_{amb}=300$\,K will produce a change in the output power of one part in 300, which corresponds to  0.1 K in the system temperature. With a system temperature of 70 K, this will produce a fractional change in the output of 0.0014. We need this to be $6.4 \times 10^{-6}$. This implies that the temperature of the lossy components should be stable to better than 0.0046 K in order for the receiver output to remain thermal noise dominated on a 100 s timescale.

\subsection{Beam calibration}
We have calculated that there are 28 sources stronger than $2$\,Jy in the region to be surveyed by BINGO. We will use their flux densities interpolated from a combination of different surveys to  establish the BINGO flux density scale. Once we determine the gain of each horn-receiver combination from measurements, we can then convert the flux density scale into a temperature scale. 

Astronomical sources, including the Moon, will pass through some of the beams and these can be used for a relatively crude calibration of the shape of the main beam. We can use the Moon since its brightness temperature as a function of lunar phase is known and it will pass through some of the BINGO beams. Also, we know its angular extent, and can deconvolve it as needed. To obtain information about the  structure of the beams' sidelobes, however, a high signal-to-noise is required.

Of the astronomical sources available, the Sun is by far the strongest and will pass at different times of the year through most of the beams. Since the radio emission from the Sun is not constant, it cannot be used for amplitude calibration, but the quiet Sun is stable enough on short timescales to measure beam shapes. This, however, is a mixed blessing since, if the solar emission can be detected in far out sidelobes, then it will be a source of an unwanted, time variable, signal. As part of the BINGO observing strategy, observations made when the Sun is close to the telescope pointing direction may have to be discarded.

On-off measurements of beam shapes are planned using a drone or a light aircraft. This is now an established technique \citep{Chang:2015}. One limitation is that the Rayleigh distance is several km. Therefore the transmitting device will need to be at a height greater than this or near-field corrections will be required.

\begin{figure*}[t]
    \centering
    \includegraphics[width=0.98\textwidth]{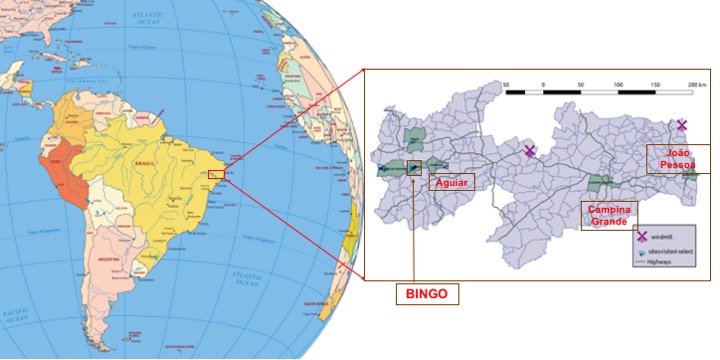}
    \caption{
    Map of BINGO location. 
    Left: Para\'\i ba state, marked by the red rectangle, is located on northeastern Brazilian coast. Brazil area, in yellow, is depicted on the map of South America. 
    Right: Para\'\i ba state map, with county subdivisions, with João Pessoa (the capital), Campina Grande (the headquarters of Universidade Federal de Campina Grande), and Aguiar counties highlighted.
    }
    \label{fig:brazil}
\end{figure*}

\subsection{Polarization calibration}
The receiver system is designed to use circular polarization because the Galactic foreground emission is partially linearly polarized and we want to be insensitive to that. 
A linearly polarized calibration signal transmitted from the secondary vertex can be used as a direct test of leakage of linear polarization into nominally circularly polarized outputs. 

The test will be performed by physically rotating the plane of the transmitted polarization and searching for any modulation in the receiver outputs. This should be a good test to check if the polarization response of the horn-polarizer combination behaves as expected and, as mentioned in the previous subsection, will be conducted as an on-off measurement.

\section{The site}
\label{sec:site}

The BINGO site was selected after weighting   the best combination of  factors such as isolation from urban areas, accessibility (but not an easy and direct access), suitable topography, and the radio quietness of the environment since RFI produced by human activities can be particularly damaging to the operation of the instrument.

We conducted a campaign to survey sites from northern Uruguay to northern Brazil, traveling (in a straight line) more than $3,000$ km, making RFI measurements in 15 different locations, as described in \cite{Peel:2019}. The quietest spot, with hardly any mobile or ADS-B transponder signals on site, is located near Aguiar, a small village about 5 h driving from Jo\~{a}o Pessoa, the capital of Para\'\i ba, in northeast Brazil. The Brazilian map, indicating Para\'\i ba state, João Pessoa, Campina Grande, and Aguiar, appears in Figure \ref{fig:brazil}. The coordinates of the fiducial site are (Lat: 7$^{\circ}$\, 2'\, 27.6''\, S; Lon: 38$^{\circ}$\, 16'\,  4.8''\, W). 

\begin{figure}[h]
    \centering
    \includegraphics[width=9cm]{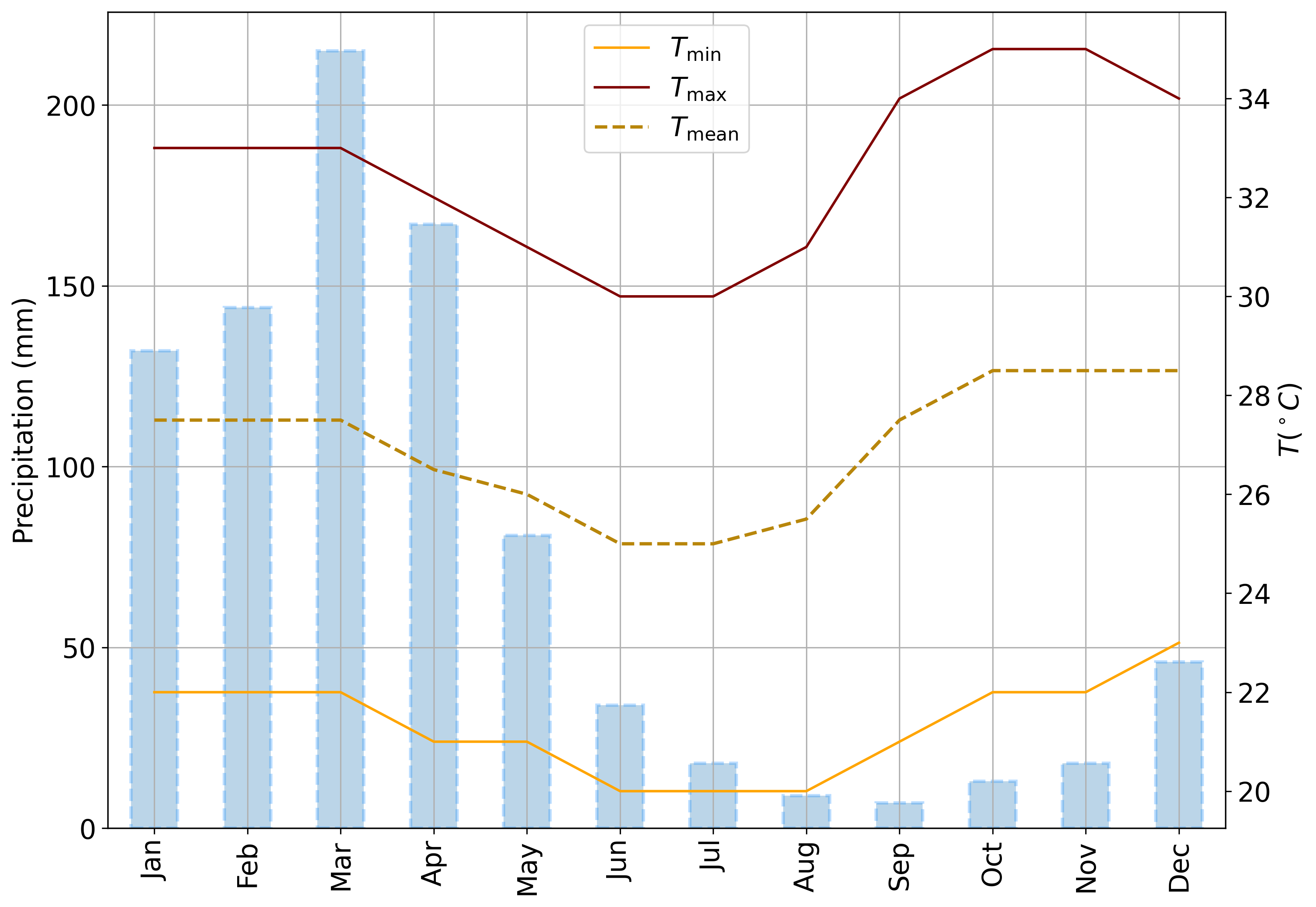}
    \caption{Temperature and precipitation variations at the Aguiar region in 2019. Climatological data are averaged over 30 years. }    
    \label{fig:clima}
\end{figure}

The typical average temperature fluctuation throughout the year in Aguiar is around $15^{\circ}$ C, with a maximum of $35.2^{\circ}$ C (in November and December) and a minimum of $19.1^{\circ}$ C (in July and August) \citep{AtmosAguiar-1,AtmosAguiar-2}. The average precipitation is $\approx 75$ mm/month, with a maximum usually between March and April (higher than $150$ mm/month) and a minimum usually lower than $10$ mm/month, between August and October. It should also be mentioned that, on top of the climatological data averaged over 30 years, there is a drought cycle correlated with the south Atlantic temperature anomaly, for roughly 4 to 5 years, whose net effect is usually to cut the average precipitation in half \citep{AtmosAguiar-3}. Figure~\ref{fig:clima} displays values for temperature and precipitation for the year 2019. 

Wind conditions are also important when considering the structural aspects of the telescope and the deformation tolerances of the reflectors. A thorough study of wind regimes in Para\'\i ba was recently completed, indicating that the BINGO region is expected to have typical winds from $5$ m/s to $7$ m/s at $150$ m height (and the wind speed is slower for smaller heights) \citep{CAMARGO-SCHUBERT:2017}.

Despite the selected site being in a very remote rural area, with a very low level of RFI contamination, as verified by a number of measurement campaigns, the BINGO band is not reserved for radio astronomy observations.\ Thus the band itself and its surroundings are subject to other RFI sources coming from the sky, the most common being radionavigation transmitters and harmonics from geostationary satellite broadcasting \citep{Peel:2019}.  

\cite{Harper:2018b} have investigated the impact of geostationary satellites in future single-dish \textsc{Hi} IM surveys, using a model of the SKA phase 1 array. They show that the GNSS signals will exceed the expected \hi\, signal of any total power IM \hi\, surveys, observing in frequencies above $950$\,MHz and in the declination range $-10 \le DEC \le -40$. One of their suggestions to suppress the GNSS signal is to build an instrument with a very good beam sidelobe suppression, which is one of the most important design requirements for BINGO. 

In an additional attempt to keep the BINGO site an RFI-free area, a request for a "radio quiet" zone around the site was requested to the Brazilian regulator agency for telecommunications (ANATEL). We consider this action an essential step to increase the chances of making high signal-to-noise observations of the desired \textsc{Hi} signal. Following the best practices adopted by other large radio observatories, the proposal considers four zones with different radii, with requirements being more strict closer to the telescope, as described below.

For Zone 1 (0.5\,km radius), radiation should be from telescope instrumentation only. Local electronic equipments should be encapsulated and communication should be made through wired devices. For Zone 2 (2\,km radius), industrial activities, power plants (wind or solar), radios, wireless networks and radio repeaters will not be allowed. For Zone 3 (10\,km radius), no new base radio stations or wind power plant will be allowed (there is only one now). For Zone 4 (25\,km radius), current mobile antennas will be kept and new antennas can be installed upon the approval of both ANATEL and the consortium. The construction of wind power plants should not be allowed in the field of view of the telescope for at least 300 km.

\section{Final considerations} 
\label{sec:status}

This paper presents the current status of the BINGO telescope, describing its optical design, receiver, calibration, and observation strategy, consisting of what we call "the fiducial BINGO." We also describe the site, updating the site selection information report in \cite{Peel:2019}. In this series of BINGO papers, besides this paper, we present a general overview of the project \citep{2020_project}, the optical system model \citep{2020_optical_design}, which was carried using the characterization of the horn prototype \citep{Wuensche:2020}, the mission simulations with the parameters listed in Table \ref{tab:instrument} \citep{2020_sky_simulation}, and two theoretical papers, one describing $21$-cm mock catalogs \citep{2020_mock_simulations} and the other presenting cosmological forecasts for BINGO \citep{2020_forecast}. 

At the time of writing this paper, the engineering projects of the dishes and the structures to support dishes and horns have been completed, and the current status of the BINGO project is summarized in the points below: 

\begin{enumerate}
        \item The horn and front end (including polarizer, transitions,  "magic tees," and rectangular-to-coaxial transitions) prototypes have been built, tested, and approved, meeting the electromagnetic and mechanical requirements;
        \item The construction of the $28$ feed horns to be used in Phase 1 has started;
        \item A prototype receiver chain has passed the initial tests;
        \item SKARABs were chosen to be our operational digital backend. The results of its integration to the BINGO receiver system will be reported in a future work;
        \item Optical design is finished, with the final dish parameters and focal plane arrangement having been simulated and tested in the mission pipeline data reduction software;
        \item The engineering project for the telescope structure has been completed;
        \item Site infrastructure (road work and terrain preparation) started in January 2021; and
        \item Construction of the dishes and structures is expected to start in the second semester of 2021. 
\end{enumerate}

\begin{figure}[h]
    \centering
    \includegraphics[width=9cm]{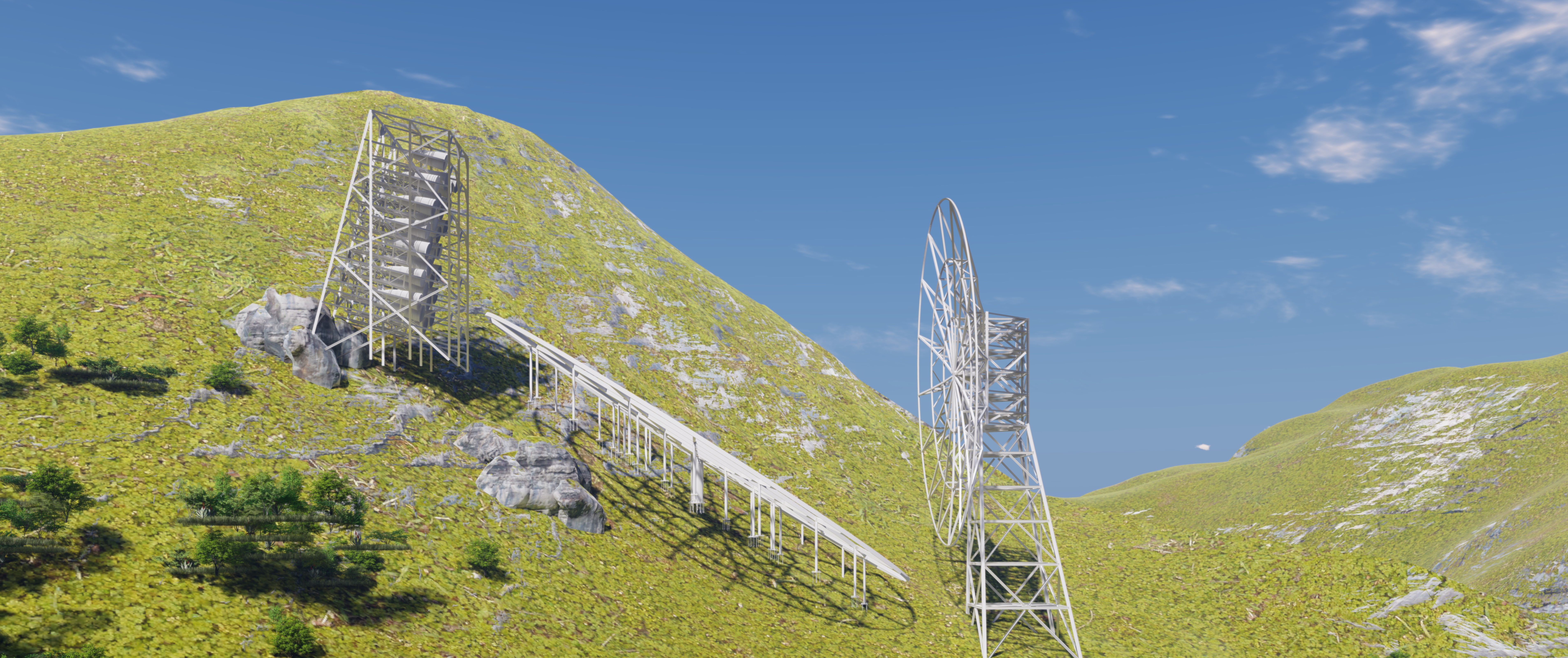}
    \caption{BINGO concept view. The optics alignment follows a north-south line, with the primary reflector pointing north. The control cabin is fully isolated electromagnetically and it is located behind the hill on the left.}
    \label{fig:architeture}
\end{figure}

The artist's view of the telescope positioned in the site is shown in Figure \ref{fig:architeture}. The view is based upon the engineering design (Figure \ref{fig:engineeringview}) and the architectonic project, with topographic curves (Figure \ref{fig:aerealview}). 

The main improvements considered for BINGO Phase 2 are the increase in the number of feed horns and upgrades in the receiver system. The optical project presented in \cite{2020_optical_design} was carried out for 28 horns, but the chosen focal plane array allows for a mechanical extension to accommodate 56 horns, doubling the redundancy in the area covered and increasing the sensitivity by $\sqrt{2}$ per year. As better LNAs and filters become available, the receivers will be upgraded for Phase 2, aiming at a smaller $T_{\textrm{sys}}$. 

We expect to see the BINGO first light by the end of 2022. If performing according to the results from the simulations and prototype tests, it will deliver high-quality \textsc{Hi} and Galactic data, as well as detect many transient phenomena. During Phase 1, BINGO will give an important contribution to the IM community, paving the way for upcoming instruments and, eventually, it will be able to measure the BAO signal in the redshift range $0.13 < z < 0.45$, contributing to the efforts of a better understanding of the nature of DE.

\begin{acknowledgements}
The BINGO project is supported by FAPESP grant 2014/07885-0. 
C.A.W. acknowledges a CNPq grant 313597/2014-6 and thanks J. W. Vilas-Boas for the useful comments and suggestions to improve the manuscript. 
T.V. acknowledges  CNPq  Grant  308876/2014-8. 
E.A. gracefully acknowledges the CNPq support. 
V.L. thanks S{\~a}o Paulo Research Foundation (FAPESP) for financial support through grant 2018/02026-0.
F.V. and E.M. acknowledge CAPES for the Ph.D. fellowships. 
M.P. acknowledges funding from a FAPESP Young Investigator fellowship, grant 2015/19936-1.
A.R.Q., F.A.B., L.B., J.R.L.S., and M.V.S. acknowledge PRONEX/CNPq/FAPESQ-PB (Grant no. 165/2018).
F.B.A. acknowledges the UKRI-FAPESP grant 2019/05687-0, and FAPESP and USP for Visiting Professor Fellowships where this work has been developed.
A.A.C. acknowledges financial support from the China Postdoctoral Science Foundation, grant number 2020M671611.
R.G.L. thanks CAPES (process 88881.162206/2017-01) and the Alexander von Humboldt Foundation for the financial support. 
K.S.F.F. thanks S{\~a}o Paulo Research Foundation (FAPESP) for financial support through grant 2017/21570-0.
C.P.N. thanks S{\~a}o Paulo Research Foundation (FAPESP) for financial support through grant 2019/06040-0.
L.S. is supported by the National Key R\&D Program of China (2020YFC2201600).
J.Z. was supported by IBS under the project code, IBS-R018-D1. 
Y.Z.M. acknowledges the support from NRF-120385, NRF-120378 and NRF-109577.
J.R.L.S. thanks CNPq (Grant no. 420479/2018-0) and CAPES for the financial support. 
M.R. acknowledges funding from the European Research Council Grant CMBSPEC (No. 725456).
The authors thank the HEALPix creators for the HEALPix package (http://healpix.sourceforge.net) \citep{Healpix:2005}.
\end{acknowledgements}

\newpage

\bibliographystyle{aa}
\bibliography{references} 

\end{document}